\documentclass[sn-mathphys-num]{sn-jnl}

\usepackage{graphicx}%
\usepackage{multirow}%
\usepackage{amsmath,amssymb,amsfonts}%
\usepackage{amsthm}%
\usepackage{mathrsfs}%
\usepackage[title]{appendix}%
\usepackage{xcolor}%
\usepackage{textcomp}%
\usepackage{manyfoot}%
\usepackage{booktabs}%
\usepackage{algorithm}%
\usepackage{algorithmicx}%
\usepackage{algpseudocode}%
\usepackage{listings}%
\usepackage{lineno}

\theoremstyle{thmstyleone}%

\usepackage{svg}
\svgsetup{
    inkscapeexe = your_path/inkscape.exe, 
    inkscapelatex = false                 
}
\theoremstyle{thmstyletwo}%

\theoremstyle{thmstylethree}%

\begin{document}

\title{Beyond Terabit/s Integrated Neuromorphic Photonic Processor for DSP-Free Optical Interconnects}


\author[1]{\fnm{Benshan} \sur{Wang}}\email{bswang@link.cuhk.edu.hk}
\author[1]{\fnm{Qiarong} \sur{Xiao}}\email{qrxiao@link.cuhk.edu.hk}
\author[1]{\fnm{Tengji} \sur{Xu}}\email{tengjixu@link.cuhk.edu.hk}
\author[1]{\fnm{Li} \sur{Fan}}\email{lifan210@link.cuhk.edu.hk}
\author[1]{\fnm{Shaojie} \sur{Liu}}\email{shaojieliu@link.cuhk.edu.hk}

\author[2]{\fnm{Jianji} \sur{Dong}}\email{jjdong@mail.hust.edu.cn}
\author[3]{\fnm{Junwen} \sur{Zhang}}\email{junwenzhang@fudan.edu.cn}
\author*[1]{\fnm{Chaoran} \sur{Huang}}\email{crhuang@ee.cuhk.edu.hk}

\affil*[1]{\orgdiv{Department of Electronic Engineering}, \orgname{The Chinese University of Hong Kong}, \orgaddress{\city{Shatin}, \state{Hong Kong SAR}, \country{China}}}

\affil[2]{\orgdiv{Wuhan National Laboratory for Optoelectronics}, \orgname{Huazhong University of Science and Technology}, \orgaddress{\city{Wuhan}, \state{Hubei}, \country{China}}}

\affil[3]{\orgdiv{School of Information Science and Technology}, \orgname{Fudan University}, \orgaddress{\state{Shanghai}, \country{China}}}



\abstract{The rapid expansion of generative AI is driving unprecedented demands for high-performance computing. Training large-scale AI models now requires vast interconnected GPU clusters across multiple data centers. Multi-scale AI training and inference demand uniform, ultra-low latency, and energy-efficient links to enable massive GPUs to function as a single cohesive unit. However, traditional electrical and optical interconnects, which rely on conventional digital signal processors (DSPs) for signal distortion compensation, are increasingly unable to meet these stringent requirements. To overcome these limitations, we present an integrated neuromorphic optical signal processor (OSP) that leverages deep reservoir computing and achieves DSP-free, all-optical, real-time processing. Experimentally, our OSP achieves a 100 Gbaud PAM4 per lane, 1.6 Tbit/s data center interconnect over a 5 km optical fiber in the C-band  (equivalent to over 80 km optical fiber in the O-band), far exceeding the reach of state-of-the-art DSP solutions, which are fundamentally constrained by chromatic dispersion in IMDD systems. Simultaneously, it delivers a four-orders-of-magnitude reduction in processing latency and a three-orders-of-magnitude reduction in energy consumption. Unlike DSPs, which introduce increased latency at high data rates, our OSP maintains consistent, ultra-low latency regardless of data rate scaling, making it an ideal solution for future optical interconnects. Moreover, the OSP retains full optical field information for better impairment compensation and adapts to various modulation formats, data rates, and wavelengths. Fabricated using a mature silicon photonic process, the OSP can be monolithically integrated with silicon photonic transceivers, enhancing the compactness and reliability of all-optical interconnects. This research provides a highly scalable, energy-efficient, and high-speed solution, paving the way for next-generation AI infrastructure.}




\maketitle

\section{Introduction}\label{sec1}

The surging demand for artificial intelligence and machine learning (AI/ML), especially in generative AI, has driven unprecedented requirements for high-performance computing infrastructure. Modern generative AI models, such as GPT-4 with its trillion-plus parameters~\cite{openai2024gpt4}, require that vast clusters of GPUs work together seamlessly, not only within a single data center but increasingly across multiple data center campuses~\cite{semianalysis2024,Zhao24ECOC}. Cross-datacenter training becomes essential to handle the sheer scale of large AI models, yet it introduces significant challenges: many GPUs must essentially function as one giant GPU. Ultra-low and uniform latency becomes critical to ensure that data is exchanged efficiently between all sites~\cite{aryalab}. As major cloud providers have pushed them to expand their AI training clusters to over 300,000 GPUs, slight latency variations will leave substantial computational resources idle, driving up operational costs~\cite{semianalysis2024}. Compounding the latency challenge is the surging energy consumption associated with data interconnect. GPU-to-GPU links consume roughly 30 picojoules per bit, and current networking costs around \$1-\$2 per Gbps—a figure that needs to be reduced by approximately 10$\times$ for cost-effective AI deployments~\cite{aryalab}. Consequently, the sustainable evolution of computing infrastructure now hinges on developing innovative data interconnect solutions that deliver consistent, ultra-low-latency, and ultra-low-energy data transfers across multiple data centers.

Although optical interconnects have been widely adopted to enable high-bandwidth, low-power connectivity over longer distances in a compact form factor~\cite{ethernetalliance2024}, they still lag behind in delivering the consistent, ultralow latency and ultralow-energy consumption required for scaling AI systems crossing multiple data centers~\cite{Zhao24ECOC}. This shortfall is largely due to limitations in digital signal processors (DSPs) required for signal distortion compensation, which, like electrical interconnects, face increased latency and energy costs at higher data rates~\cite{aryalab}. Intensity Modulation/Direct Detection (IMDD) is widely used for data center interconnects (DCIs) because of its simplicity, low cost, and energy efficiency. However, its baud rate and distance scalability are severely constrained by chromatic dispersion~\cite{chagnon2019,che23JLT}. This limitation arises because phase information is lost during direct detection, preventing DSPs from fully compensating for chromatic dispersion. As a result, even advanced DSP chips such as Marvell’s Ara 1.6T PAM4 DSP engine~\cite{marvell24ara}, built on a 3 nm CMOS process for 8$\times$200G Ethernet, can only extend the reach of 112 Gbaud PAM4 signals to about 2 km at the edge wavelengths of the O-band, where chromatic dispersion is already minimal~\cite{liu22,bernal24NC}. Moreover, efforts to increase channel numbers by reducing channel spacing introduce significant nonlinearity penalties~\cite{berikaa23JLT}.


Neuromorphic optical processors have been proposed as a promising alternative to DSPs~\cite{shastri21NP,huang22APX,zhou22LSA,markovic20NRP,fu24LSA,mcmahon23NRL,chen23Nature,bandyopadhyay24NP,xu21Nature,ashtiani22Nature,sludds22Science}, with demonstrations in both IMDD~\cite{argyris18SR,argyris19Access,sackesyn21OE,Wang22JSTQE,staffoli23PR,shen23Optica,gooskens23SR,staffoli24JLT,sozos24JLT} and coherent~\cite{huang21NE,masaad23Nano,masaad24JLT,sozos22CP} transmission systems. However existing systems have yet to deliver capabilities approaching existing DSP chips or achieve the processing speeds and bandwidth that optical processors are theoretically capable of offering (Fig.~\ref{fig:Fig2_Literature}a). Experimental demonstrations have so far achieved a maximum speed of only 40 Gbaud/$\lambda$~\cite{liu24NC}, and numerical simulations reach 60 Gbaud/$\lambda$~\cite{liu24OFC}. The speed is constrained either by the photonic circuits~\cite{argyris18SR,argyris19Access,sackesyn21OE, shen23Optica,staffoli23PR,gooskens23SR,staffoli24JLT} or optoelectronic devices providing nonlinear activation functions (optical neurons)~\cite{Wang22JSTQE,huang21NE}. Furthermore, while some approaches realize complex communication systems, they rely heavily on pre- or post-processing in the digital domain, introducing significant overhead from domain conversions that negate the efficiency gains of optical processing~\cite{argyris18SR,argyris19Access,shen23Optica,masaad23Nano,masaad24JLT,sozos24JLT,sozos22CP}. In systems that leverage reservoir computing~\cite{yan24NC}, large readout layers are required and are all implemented digitally, requiring multiple high-speed ADCs matching the speed of optical communication signals~\cite{argyris18SR,argyris19Access,shen23Optica,masaad23Nano,masaad24JLT}. This results in unsustainable energy consumption and processing latency~\cite{ludge24Nature}. Moreover, most photonic processors are limited to fixed data rates, modulation formats, or single wavelengths, lacking the adaptability and programmability needed to support varying link conditions~\cite{sackesyn21OE,staffoli23PR,gooskens23SR,masaad23Nano,masaad24JLT}. In summary, current optical processors have yet to deliver capabilities approaching those of DSPs.

\begin{figure}[htbp] 
    \centering 
    \includegraphics[width=1\textwidth]{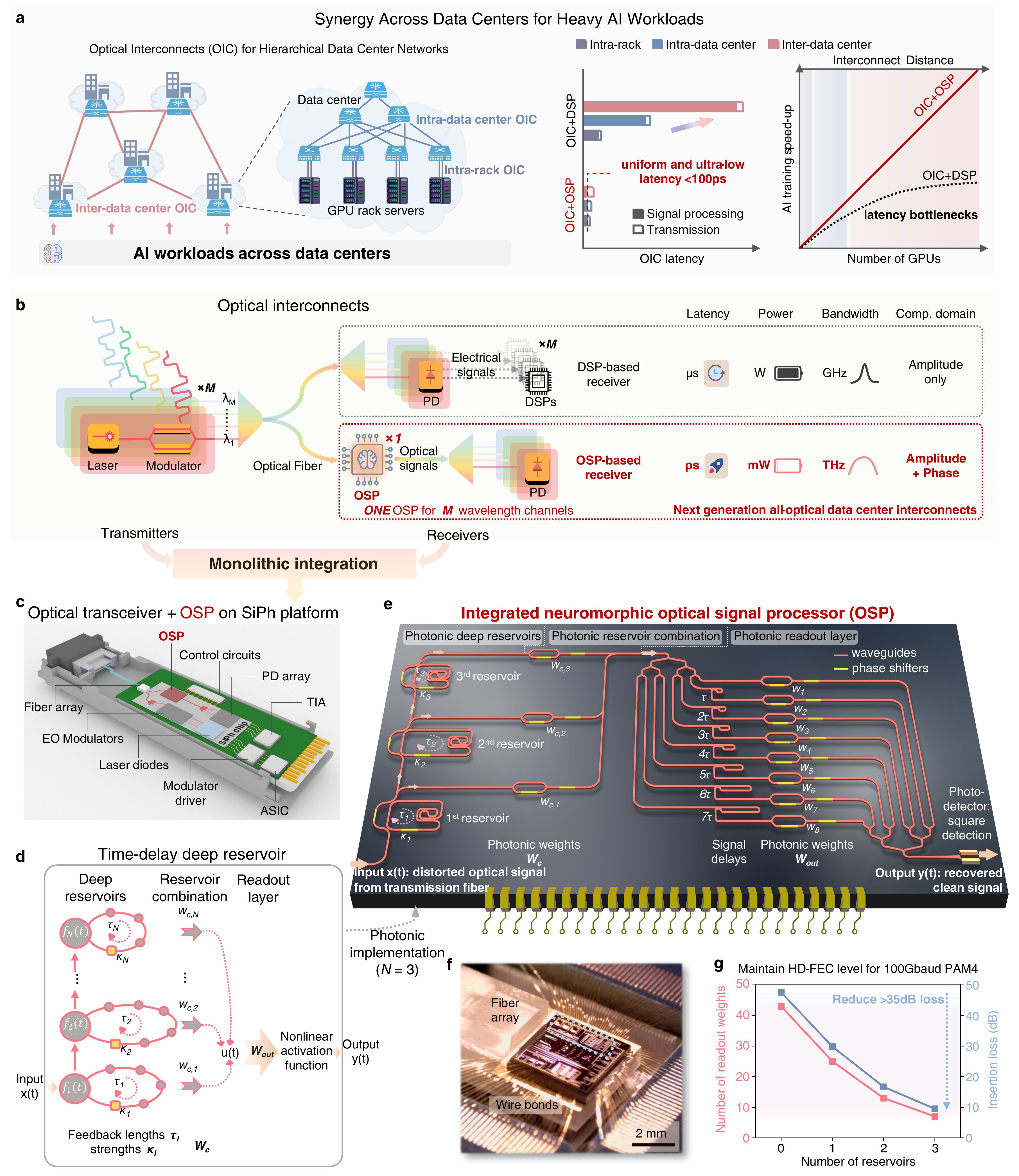} 
    \caption{OSP architecture and implementation. (a) Multi-data center interconnection for AI/ML workloads. Optical interconnects (OIC) enabled by OSP achieve ultra-low and uniform latency, even as the scale of interconnected data centers grows. OSP overcomes the latency bottlenecks traditionally caused by DSPs, enabling accelerated AI training across increasingly large and multi-scale GPU clusters. (b) Optical interconnects using IMDD schemes with DSP-based and OSP-based receivers. One single OSP can support $M$ wavelength channels, whereas $M$ DSPs are required to handle the same number of channels. The shown photodiode (PD) represents the receiver front-end, which also includes the trans-impedance amplifier (TIA). Comp.: compensation. (c) Optical module integrating photonic components (OSP, PDs, electro-optic (EO) modulators, and laser diodes) with electrical components (ASIC, modulator driver, TIA, and control circuits). Optical transceiver and OSP are integrated on a single silicon photonic chip. This compact design enables high-performance optical signal processing in a scalable form factor. (d) Time-delay deep reservoir scheme, consisting of $N$ reservoir nodes with different delayed self-feedback lengths $\tau_l$ and feedback strength $\kappa_l$. Dynamical states of different reservoirs are combined using $W_c$ and then processed through a readout layer $W_{out}$ to generate the final output. (e) Proposed OSP architecture comprising three photonic reservoirs and eight readout channels. (f) The packaged OSP chip with electrical wire bonds and an optical fiber array. (g) Performance comparison of different reservoir numbers to maintain the hard-decision forward error correlation (HD-FEC) level for 100 Gbaud PAM4 signals over 5 km of fiber transmission in the C-band.} 
    \label{fig:Fig1_architecture} 
\end{figure}

\begin{figure}[htbp] 
    \centering 
    \includegraphics[width=1\textwidth]{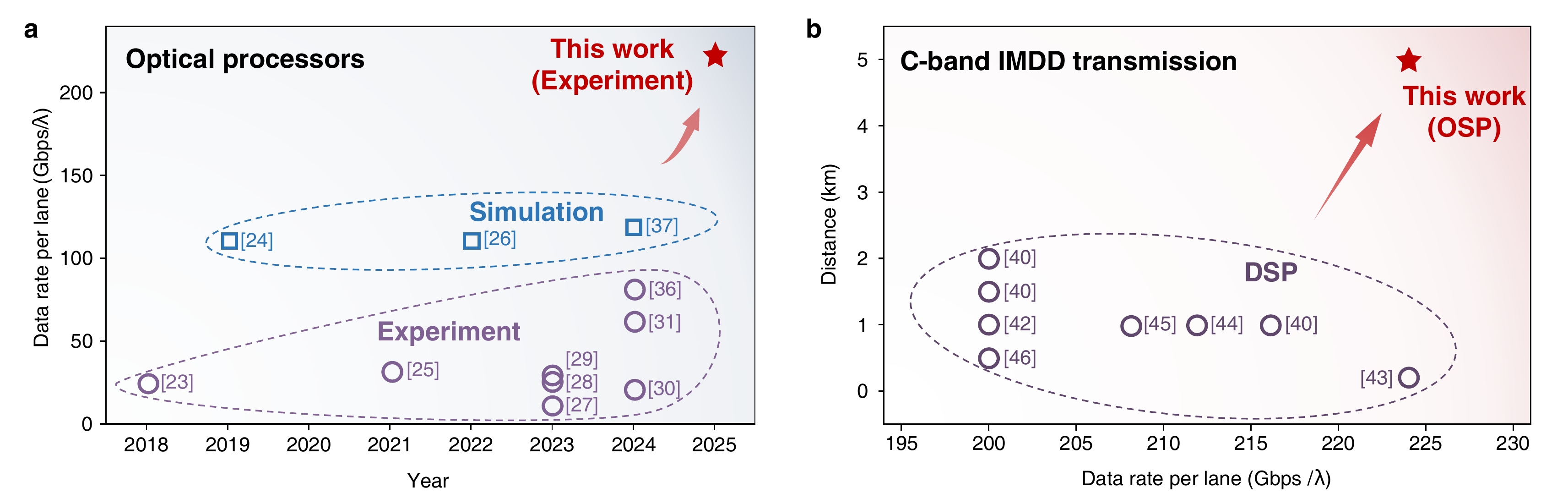} 
    \caption{Comparison with published works. (a) Comparison of processing data rate per lane between our work and other optical processors~\cite{argyris18SR,argyris19Access,sackesyn21OE,Wang22JSTQE,staffoli23PR,shen23Optica,gooskens23SR,staffoli24JLT,liu24NC,liu24OFC, sozos24JLT}. (b) Comparison of beyond 200 Gbps/$\lambda$ C-band standard IMDD fiber transmission using single PD between our OSP and other DSP methods~\cite{chan22JLT,chan22OL,cheng23OL,han23OL,sang22JLT,verbist19JLT,mardoyan17JLT}.} 
    \label{fig:Fig2_Literature} 
\end{figure}

In contrast, this work presents the first experimental demonstration of an integrated OSP enabling all-optical, DSP-free, real-time processing for 1.6 Tbit/s DCIs, with signal compensation performances surpassing DSP. Our OSP experimentally delivers ultra-low latency ($<$60 ps) and ultra-low energy consumption (fJs/bit), while supporting high-speed operation ($>$100 Gbaud/$\lambda$) and multi-channel performance (eight 100 Gbaud PAM4 channels at 1.6 Tbit/s). Compared to 3 nm CMOS-based DSP solutions, it delivers over 3-order-of-magnitude improvements in both latency and energy efficiency. By leveraging the vast bandwidth of the lightwave, the OSP can easily scale to even higher data rates per channel and support higher channel counts without incurring significant increases in latency or energy consumption. Moreover, it outperforms state-of-the-art DSP hardware and algorithms by enabling 10 times longer transmission distances at speeds exceeding 100 Gbaud, as summarized in Fig~\ref{fig:Fig2_Literature}~\cite{chan22JLT,chan22OL,cheng23OL,han23OL,sang22JLT,verbist19JLT,mardoyan17JLT}. Unlike conventional IMDD links, where phase information is lost during direct detection, our OSP retains and leverages the full optical field before signal detection, enabling advanced impairment compensation that outperforms DSPs. It successfully enables 1.6 Tbit/s DCIs over 5 km of optical fiber in the C-band. The corresponding chromatic dispersion ranges from 82 ps/nm to 90 ps/nm between 1540 nm and 1565 nm—equivalent to the dispersion experienced over 80 km of fiber in the O-band of 1300–1325 nm. In contrast, the recent 3 nm DSP chip only delivers a 2 km transmission distance in the O-band~\cite{marvell24ara}. We also demonstrate that the OSP is programmable and adaptable to different modulation formats, data rates, and wavelengths through $in\text{-}situ$ learning. 

As shown in Fig.~\ref{fig:Fig1_architecture}c, the demonstrated OSP is fabricated using a mature silicon photonic foundry, enabling monolithic integration of optical signaling (i.e., optical transceivers) and processing on the same silicon photonic chip. By delivering uniform, ultra-low latency, and energy consumption that remain constant even as data rates scale, our OSP directly tackles the critical challenges of multi-scale AI training—where trillions of parameters and vast GPU clusters demand seamless, synchronized data exchange—paving the way for the next generation of efficient, scalable, and cost-effective AI infrastructure.

\section{Results}\label{sec2}

\subsection{Principle and chip design}\label{subsec2}
Our OSP leverages deep reservoir computing (RC)—a neuromorphic approach inspired by deep recurrent neural networks. RC is a three-layer NN comprising an input layer, a reservoir layer, and a readout layer~\cite{yan24NC}. The input layer receives information and performs initial processing, the reservoir layer consists of nonlinear nodes with random, fixed connections, and the readout layer combines signals from the reservoir to generate the desired output through training. RC is particularly effective at processing temporal information. It also simplifies the training of recurrent NNs since only the readout layer requires training. 

To overcome the limitations in prior OSP systems, we focus on two primary goals: first, to ensure the capability to handle large dispersion at high speeds with minimal optical components and optical losses; and second, to design a photonic circuit that is highly scalable to support both high data rates and multiple wavelengths without constraints or overhead. To achieve these goals, our chip design introduces several innovations that distinguish it from prior works. To address large dispersion, we adopt a novel deep reservoir architecture. The deep reservoir significantly enhances memory capacity and dynamic processing capabilities compared to a single reservoir~\cite{goldmann20Chaos,gallicchio17Neuro}. As a result, the deep reservoir architecture reduces the number of required readout parameters, thereby minimizing the number of optical components and associated optical losses. Specifically, as shown in Fig.~\ref{fig:Fig1_architecture}f, adding three additional reservoirs reduces 35 readout parameters and over 35 dB insertion loss while maintaining the bit error rate (BER) level. To ensure the OSP provides sufficient bandwidth to accommodate high-speed signals and multiple wavelengths, we introduce two new designs. First, we eliminate the need for the input layer typically used in RC. This input layer, often implemented using digital masks operating at rates several times faster than the input signal, is not practical and becomes a speed and energy bottleneck when input signal speeds exceed 100 Gbaud. Second, our OSP avoids using bandwidth-constrained or wavelength-selective optoelectronic components—such as lasers~\cite{argyris18SR,shen23Optica}, modulators~\cite{lupo23Optica,picco24TNNLS}, or ring resonators~\cite{denis18JSTQE,donati21OE}, that are commonly used in photonic RC. Instead, it is constructed using optical waveguides and incorporates a single high-speed photodetector at the output to handle both signal detection and nonlinearity. The high-speed photodetector will directly generate clean and recovered signals. 

Fig.~\ref{fig:Fig1_architecture}d and~\ref{fig:Fig1_architecture}e shows the detailed OSP design. The proposed OSP is a complex-valued deep time-delay RC system. At the receiver, the distorted optical communication signal from the transmission optical fiber is optically coupled into the OSP and enters the RC layer without an input layer. The deep time-delay RC consists of $N$ nodes with states $f_l(t),l=1,2,..., N$, where each node functions as an independent reservoir and introduces a feedback delay, denoted as $\tau_l$, with a corresponding feedback strength, $\kappa_l$, as illustrated in Fig.~\ref{fig:Fig1_architecture}d. The number of reservoirs, along with the feedback delays and strengths, are hyperparameters optimized according to the specific task requirements. The output of each reservoir is weighted by $W_{c,l}$ and then combined. Deep-layered RC architecture with varying time delays effectively projects the input signals into a higher-dimensional space in the time domain. This eliminates the need for a traditional input layer. Weights $W_{c,l}$ is optimized to balance the output power of different reservoirs. The RC layer is followed by a readout layer with weights $W_{out}$. The weights in the RC are complex-valued. A nonlinear activation function is added to the readout layer output, providing the nonlinear mapping between the input and output signals

Building on the deep time-delay RC architecture, our OSP consists of a photonic deep reservoir with three cascaded reservoir nodes. The outputs of these layers are weighted, combined, and finally processed by the photonic readout layer with 8 readout channels, as shown in Fig.~\ref{fig:Fig1_architecture}e. The reservoir layer comprises three feedback loops, each having a fixed delay time and a tunable delay strength. This feedback loop generates multiple copies of the input waveform, each with a time delay determined by the loop length. The delay times in the three reservoir layers are designed to be approximately 18 ps, 36 ps, and 9 ps, respectively, designed to process signals with 100 Gbaud and beyond. These delay times are chosen to be non-integer multiples of the signal data rate to enrich the temporal dynamics. The feedback strength in each layer is tunable through a Mach-Zehnder interferometer (MZI) coupler. Additionally, a phase shifter is added within the loop to program the signal phase, making the reservoir complex-valued. The three reservoir nodes are cascaded. For the first and second nodes, their outputs are split into two branches by a 2:1 and a 1:1 coupler, respectively. One branch is directed to the subsequent reservoir node, while the other branch serves as the output and is weighted by a complex coefficient $w_{c,l}$, which is realized using an MZI in conjunction with a phase shifter. After weighting, the outputs of each reservoir layer are combined to produce the input to the photonic readout layer. (See Supplementary Note 1 for the detailed model of a single photonic reservoir). The deep reservoir layer generates an output that is a rich combination of multiple neighboring waveforms crossing a long-time span, which effectively projects the original input into a high-dimensional space in the time domain.

The photonic readout layer functions equivalently as a tapped delay filter with eight taps and complex-valued coefficients. The input to the readout layer is first split into eight copies, with each copy introducing a delay that increases from 0 to 7$\tau$, where $\tau$ is designed to be 5 ps to enable dense sampling for analog processing. Each delayed copy is weighted by a complex readout weight, which is realized using an MZI in conjunction with a phase shifter. We demonstrate in Fig.~\ref{fig:Fig1_architecture}f that the deep reservoir significantly reduces the required number of readout weights. Without the reservoir, the number of readout weights needed to bring the error below hard-decision forward error correlation (HD-FEC) threshed would increase to 43, resulting in over 35 dB additional insertion loss. The weighted delayed copies are then combined and detected by a photodetector. The photodetector performs a nonlinear transformation through the $|\cdot|^2$ operation. The output of the photodetector is the recovered signal, thus realizing an all-optical processor that operates without any assistance from DSP. 

The OSP is fabricated using a commercial silicon-on-insulator (SOI) foundry platform and can be monolithically integrated with photonic transceivers on the same silicon photonic chip. The phase shifters are implemented by microheaters (see Method for chip design and fabrication details). Through $in\text{-}situ$ training of phase shifters, the OSP can effectively compensate for both linear and nonlinear channel impairments for signals with various data rates, modulation formats, and wavelengths.

\subsection{DSP-free high-speed DCI transmission with OSP}\label{subsec2}

\begin{figure}[htbp] 
    \centering 
    \includegraphics[width=1\textwidth]{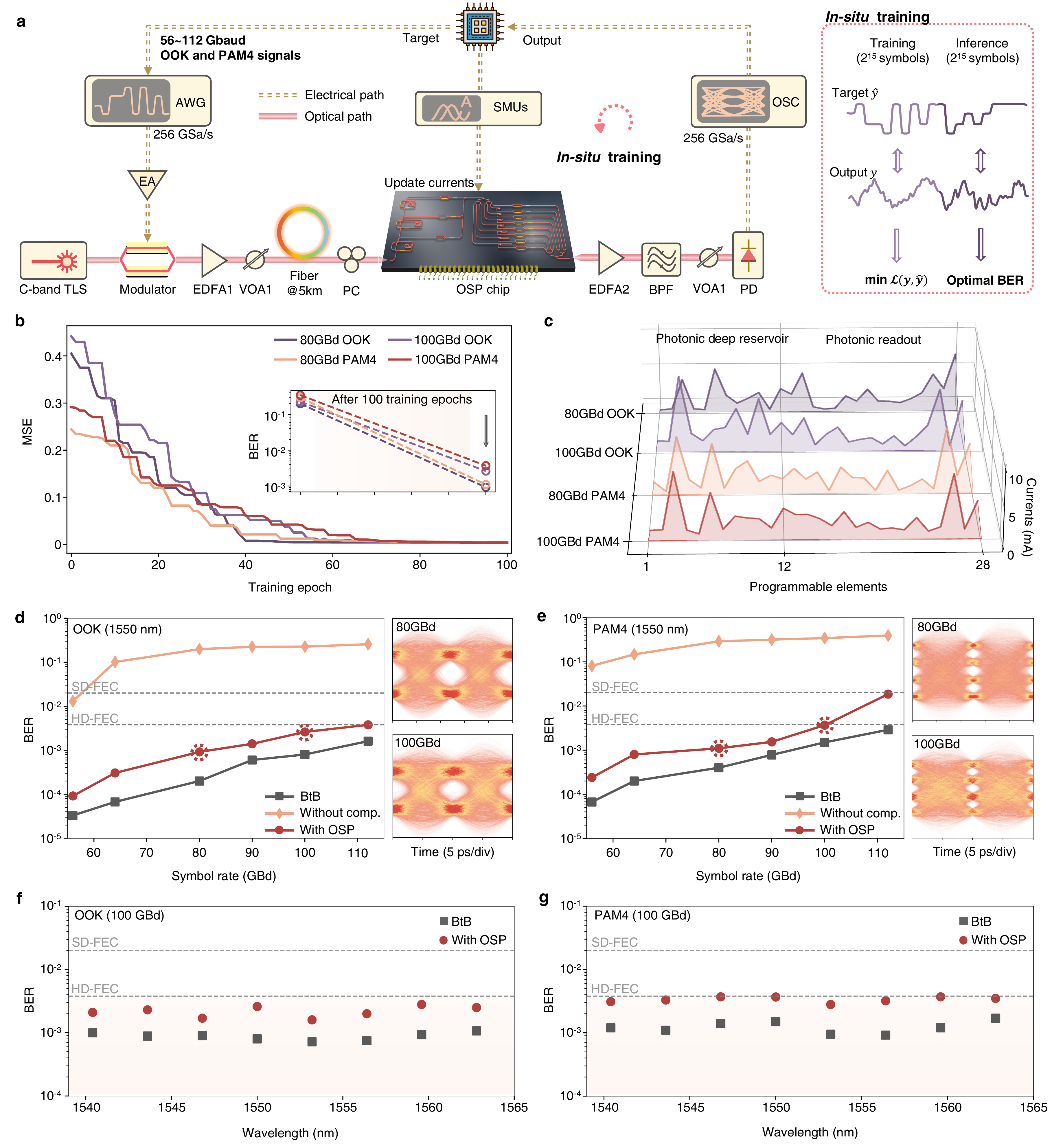} 
    \caption{High-speed signal processing and programmability. (a) Schematic of the experimental setup and training process. TLS tunable laser source, PC polarization controller, EDFA erbium-doped fiber amplifier, VOA variable optical attenuator, AWG arbitrary waveform generator, EA electrical amplifier, SMUs source meter units, BPF band-pass filter, PD photodiode, RTO real-time oscilloscope. (b) The MSE converges as the OSP is trained using $2^{15}$ symbols of OOK and PAM4 signals at 80 Gbaud and 100 Gbaud. (c) Optimized currents of OSP’s on-chip programmable elements for different symbol rates and modulation formats. (d)(e) Measured BER as a function of symbol rates for OOK and PAM4 signal transmission, horizontal dashed lines indicate the thresholds for soft-decision and hard-decision forward error correlation (FEC). The associated eye diagrams are shown for the circled points. (f)(g) Measured BER as a function of wavelength for 100 GBaud OOK and PAM4.} 
    \label{fig:Fig3_Setup} 
\end{figure}

We demonstrate the implementation of the OSP for IMDD DCIs using the experimental setup illustrated in Fig.~\ref{fig:Fig3_Setup}a. The transmission experiment is conducted over a 5-km single-mode fiber (SMF) operating at the C-band. The C-band is chosen because most equipment in our laboratory operates at the C-band. The corresponding dispersion is 85 ps/nm at 1550 nm, equivalent to over 80 km transmission distance at the O-band of 1300-1325 nm~\cite{sector16characteristics}. At the transmitter, high-speed optical signals are generated using the experimental setup described in the Method section. To minimize the use of DSP, pre-emphasis is not applied at the transmitter. A root-raised cosine (RRC) filter with a roll-off factor of 0.05 is used to balance spectral efficiency and timing jitter tolerance. At the receiver, the OSP is deployed before the photodetector to directly compensate for transmission impairments in the optical domain. The detected signal is sampled by a real-time oscilloscope and then counted the bit error at one sample per symbol without oversampling. No additional DSP is used at the receiver. 

Our OSP is reconfigurable, enabling it to adapt to dynamic transmission conditions and support various modulation formats, data rates, and wavelengths. This contrasts with other photonic processors and dispersion compensators, which are typically non-reconfigurable and remain fixed after fabrication or an initial setup. The programmability is achieved through 28 phase shifters at both the reservoir and the readout layer, each operating as an independent trainable unit. By optimizing the currents applied to these phase shifters, the OSP can approximate the inverse channel response of the DCI, effectively compensating for both linear and nonlinear impairments simultaneously.

To demonstrate the high-speed processing and programmable capabilities of our OSP, we conduct transmission experiments using OOK and PAM4 signals at symbol rates ranging from 56 to 112 Gbaud. $In\text{-}situ$ training is employed to optimize the on-chip parameters, enabling the chip to accommodate various fiber signals with different data rates, modulation formats, and wavelengths. The phase shifters are optimized based on the particle swarm optimization (PSO) algorithm to minimize BER. At the training phase, $2^{16}$ symbols are generated from a pseudo-random binary sequence. $2^{15}$ symbols are used for training, while the remaining symbols are used for validation. The training objective is to minimize the mean square error (MSE) between the received symbols $y(t)$ after the OSP and the transmitted symbols $\hat{y}(t)$, defined as $\frac{1}{M}\sum_{i=1}^{M} [y(t_i) - \hat{y}(t_i)]^2$, where $M$ is the number of training symbols, which is $2^{15}$. 

To illustrate the adaptability of our OSP to different modulation formats and data rates, Fig.~\ref{fig:Fig3_Setup}b presents the MSE reduction during training for four signals: OOK and PAM4, each at 80 Gbaud and 100 Gbaud. For example, before training, 100 Gbaud OOK and PAM4 signals have BERs beyond $10^{-1}$ level, measured at a received optical power of 6 dBm. After approximately 100 training epochs, both BERs reduce to nearly $10^{-3}$. During training, the current values are confined to a maximum of 13 mA, which is capable of providing a 2$\pi$ phase shift. The optimal current configurations for the four signals are shown in Fig.~\ref{fig:Fig3_Setup}c. After training, we evaluate the BER performance with and without our photonic processor at different data rates, as shown in Fig.~\ref{fig:Fig3_Setup}d and~\ref{fig:Fig3_Setup}e. The results demonstrate that our OSP supports up to 112 Gbaud OOK and 100 Gbaud PAM4 signals, achieving BERs below the HD-FEC threshold. Moreover, clear open eye diagrams are obtained without requiring additional DSP compensation. Even for the 112 Gbaud PAM4 signal, the BER satisfies the soft-decision forward error correction (SD-FEC) requirements. Notably, our OSP requires no additional power to support higher data rates, in stark contrast to DSP.

To illustrate the compatibility of our OSP to different wavelengths, we also separately optimize the OSP chip for different wavelengths with 400-GHz channel spacing, ranging from 1540 nm to 1565 nm. We test the performances in both back-to-back (BtB) configurations and 5-km SMF C-band transmission. The 25-nm wavelength range crossing the C-band showcases the broad working bandwidth of the OSP, which is enabled by tunable operation through integrated microheaters and the ultra-wideband edge coupler. Fig.~\ref{fig:Fig3_Setup}f and~\ref{fig:Fig3_Setup}g present the measured BERs for BtB configuration (black) and 5 km optical fiber transmission with OSP compensation (red) for 100 Gbaud OOK and PAM4 signals, respectively. The BtB performance confirms consistent signal quality across all eight wavelengths at the transmitter. At the receiver, although fiber dispersion increases at longer wavelengths, OSP effectively compensates for signal distortions across all channels, ensuring that all BERs meet HD-FEC requirements for both 100 Gbaud OOK and PAM4 signals.

\subsection{Capability for compensating linear and nonlinear impairments}\label{subsec2}

\begin{figure}[htbp] 
    \centering 
    \includegraphics[width=1\textwidth]{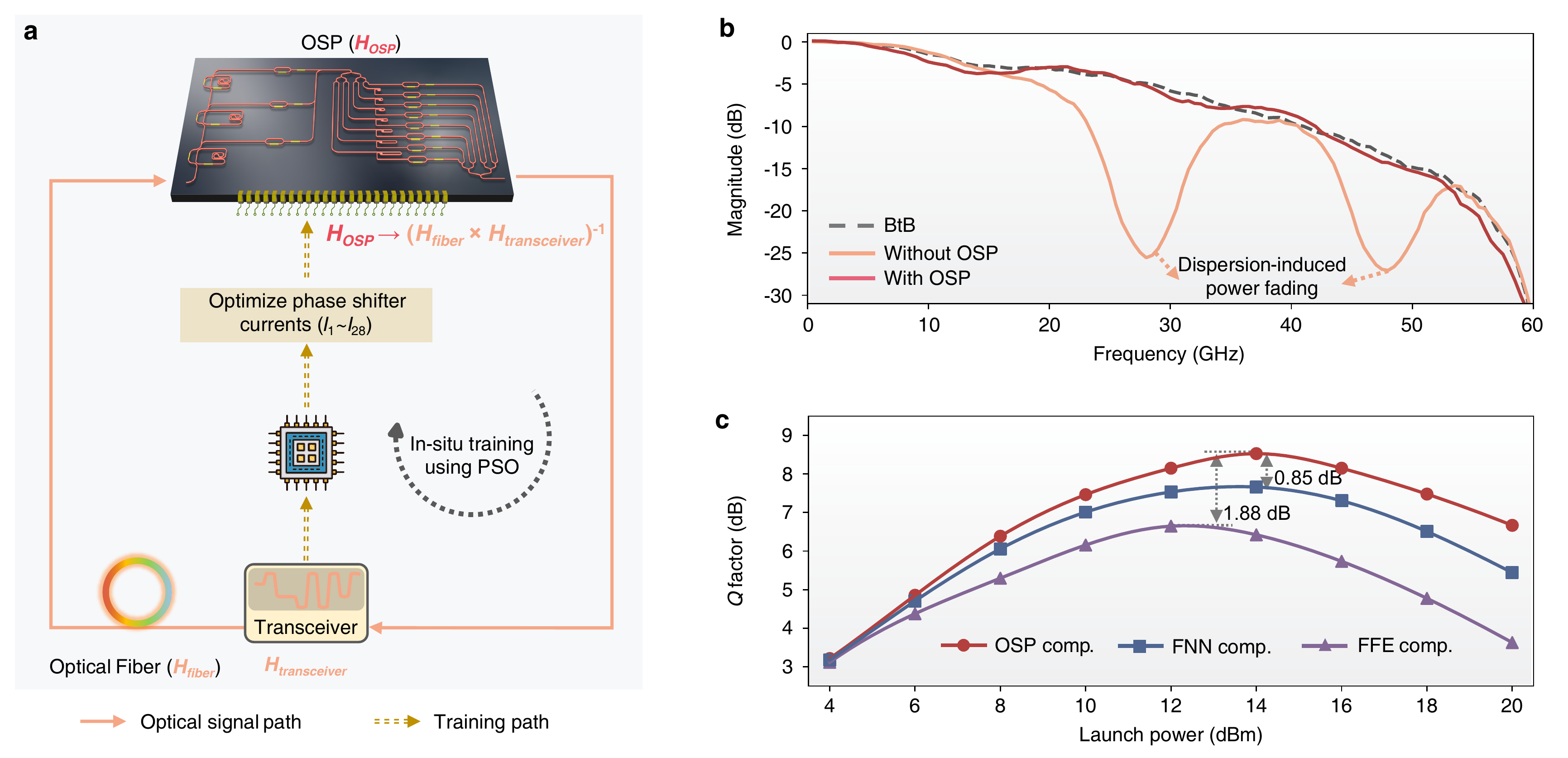} 
    \caption{Linear and nonlinear compensation. (a) Flowchart of the OSP optimization by learning the inverse complex-valued transfer function of the optical fiber and transceiver. (b)The spectral responses after detection of different situations. The dispersion-induced power fading effect, due to 5 km C-band transmission, is observed with multiple spectral nulls. With the proper optimization of our OSP, we can almost eliminate power fading caused by the fiber chromatic dispersion. (c) $Q$ factor versus launch power for 100-Gbaud PAM4 signal compensation using our OSP and other DSP methods.} 
    \label{fig:Fig4_LANL} 
\end{figure}

We provide a more detailed analysis of our OSP's capabilities, highlighting its effectiveness in compensating for both linear and nonlinear impairments in DCI systems. Typical DSP methods compensate for linear and nonlinear impairments using separate DSP modules, and the phase information is lost after direct detection. In contrast, our OSP can assess the full optical field. Meanwhile, OSP treats the overall response of the transmission system as a whole, as illustrated in Fig.~\ref{fig:Fig4_LANL}a. It adjusts its trainable parameters to approximate the inverse transfer response of the entire system, enabling compensation for various linear and nonlinear impairments within a single module while simultaneously achieving better performance than DSP.

We first examine its ability to compensate for chromatic dispersion, which is the primary source of linear distortion in high-speed DCI applications. Chromatic dispersion introduces frequency-dependent phase shift~\cite{chagnon2019}, denoted as $H_{\text{fiber}} = e^{-j \frac{\beta_2 L}{2} \omega^2}$. After square-law intensity detection, this dispersion leads to frequency-selective spectral fading, which significantly degrades signal integrity. While traditional DSP techniques can mitigate some performance penalties, they cannot fully recover the frequency loss because phase information is lost during intensity detection. In contrast, Fig.~\ref{fig:Fig4_LANL}a demonstrates that our OSP can achieve ideal chromatic dispersion compensation by learning an inverse complex-valued transfer function to \( H_{\text{fiber}} \) through parameter optimization. This is evidenced by the fact that the frequency response of the DCI channel after applying the OSP closely matches that of the BtB scenario (without an optical fiber), as shown in Fig.~\ref{fig:Fig4_LANL}b. In contrast, without the OSP, significant power fading is observed after the 5-km fiber transmission. Note that the measured frequency bandwidth is limited by the characterization equipment, not the OSP (See Supplementary Note 2 for the frequency measurement details).

The OSP is also capable of compensating for nonlinear impairments. To demonstrate this effect, we intentionally launch high optical power into the fiber to induce Kerr nonlinearity. While Kerr nonlinearity is relatively minor in single-wavelength transmission, it can become significant in multi-channel transmission, especially at the O-band. Rather than solely learning the nonlinear phase shifts induced by the Kerr effect, the OSP is trained to learn the overall inversed channel response ($H_{\text{fiber}}\times H_{\text{transceiver}}$), including both nonlinear and linear distortions in the optical domain, by minimizing the MSE between the transmitted and received symbols in the training dataset. We demonstrate the superior performance of our OSP by comparing it with typical DSP algorithms~\cite{huang22JLT}, including Feed Forward Equalizer (FFE) and Feedforward Neural Network (FNN) equalizer. We evaluate the signal $Q$ factors as a function of launched optical powers for 100 Gbaud PAM4 signal over 5 km optical fiber transmission, as shown in Fig.~\ref{fig:Fig4_LANL}c. The $Q$ factor is calculated using the formula $Q = 20 \cdot \log_{10} ( \sqrt{2} \cdot \text{erfc}^{-1}(2 \cdot \text{BER}))$~\cite{freude12ICTON} after counting the BER. FFE operating in the electrical domain applies a series of weighted taps to the received signal to mitigate linear distortions, which limits its ability to compensate for nonlinear distortions. As a result, even if employed with 885 taps, the FFE exhibits the worst BER performance, especially in the high-power region. FFN is capable of compensating for both linear and nonlinear distortions. However, because the input signal to FNN lacks phase information, even with a large FNN size (256\(\times\)256\(\times\)1), its performance remains much worse than that of OSP. Our OSP achieves the highest $Q$ factor of 8.51 dB, which is 1.88 dB higher than FFE and 0.85 dB higher than FNN. It also supports 2 dB higher launched optical power compared to FFE.

\subsection{1.6T WDM DCIs with OSP}\label{subsec2}
\begin{figure}[htbp] 
    \centering 
    \includegraphics[width=1\textwidth]{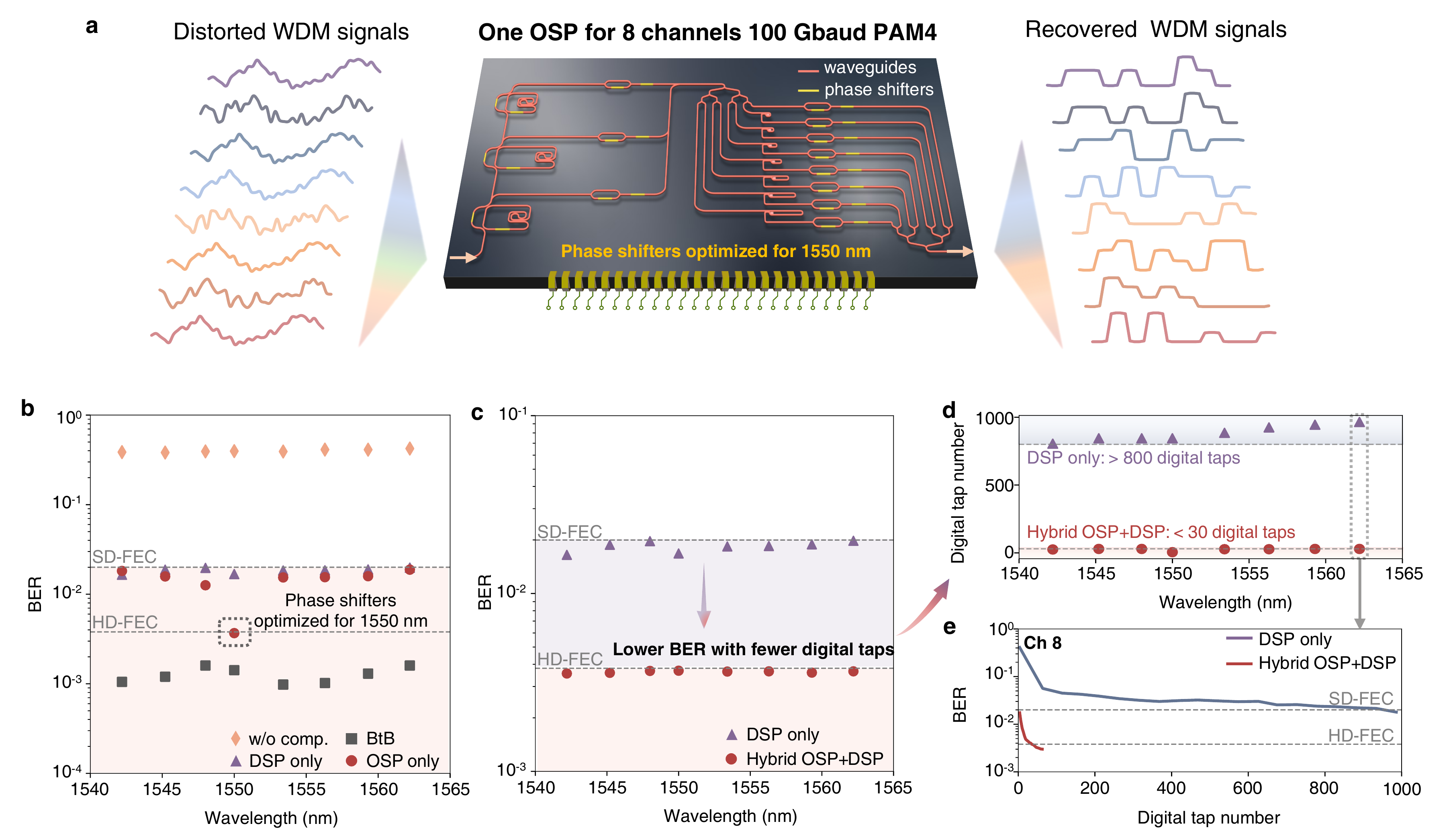} 
    \caption{1.6T WDM DCIs implementation. (a) OSP-based 1.6T silicon photonic transceiver.  (b) Measured BERs for 8 WDM channels 100 Gbaud PAM4 signals under the BtB transmission, 5 km SMF transmission without any compensation, with OSP only compensation, and with DSP only compensation. The OSP is optimized at 1550 nm (Ch 4). (c) Measured BERs for 8 WDM channels 100 Gbaud PAM4 signal under 5 km SMF transmission with hybrid OSP/DSP compensation and 5 km SMF transmission with DSP only compensation. (d) Required number of DSP taps for OSP and DSP compensation to achieve corresponding BERs in (c). (e) BER versus the number of DSP taps for OSP and DSP compensation at 1562.2 nm (Ch 8). } 
    \label{fig:Fig5_1.6T transceiver} 
\end{figure}

To support high-throughput DCIs, WDM-based DCIs are essential. The number of DSP modules, along with their associated power consumption, footprint, and thermal management challenges, all scale with the number of wavelengths. In contrast, by leveraging the inherent parallelism and large bandwidth of lightwave, processing multiple wavelength channels only requires one OSP without incurring additional costs such as energy consumption or increasing chip footprint, as shown in Fig.~\ref{fig:Fig5_1.6T transceiver}a. 

To demonstrate this advantage, we implement the OSP in a 1.6 Tb/s WDM DCI system with an eight-channel WDM transmission over a 5 km SMF link in the C-band, where each channel carries 200 Gbps PAM4 signals. The transmitted wavelengths span from 1540 nm to 1565 nm, with a channel spacing of approximately 400 GHz.

Since different channels have different dispersions, here we discuss two potential implementations. In the first implementation, the OSP is initially trained and optimized at 1550 nm. Once trained, the OSP is applied across all channels. Different from DSP, our OSP simultaneously processes multiple wavelength channels without incurring additional energy consumption or increasing chip footprint. As shown in Fig.~\ref{fig:Fig5_1.6T transceiver}b, since OSP is only optimized for one channel, wavelength-dependent dispersion introduces residual impairments in non-optimized channels, resulting in an increased BER. However, it is important to note that even if residual impairment remains uncompensated in OSP, the BER performances of OSP for all channels are still comparable to or even outperforms DSP, even though DSP is independently optimized for each channel. In addition, the BERs of all channels are lower than the SD-FEC threshold.

In the second implementation, we add DSP after the OSP to compensate for residual impairments. Since the major impairments have already been addressed by the OSP, the DSP modules are very lightweight. In this case, the BERs of all WDM channels successfully fall below the HD-FEC threshold, as shown in Fig.~\ref{fig:Fig5_1.6T transceiver}c. Fig.~\ref{fig:Fig5_1.6T transceiver}c compares the BERs of our hybrid OSP/DSP scheme and DSP-only scheme. Even with more than 800 taps for each channel, DSP alone can only achieve the SD-FEC threshold. In stark contrast, the hybrid OSP/DSP scheme enables performances below the HD-FEC threshold, and the post-DSP is very lightweight, requiring less than 30 taps for each channel. For instance, as depicted in Fig.~\ref{fig:Fig5_1.6T transceiver}d and~\ref{fig:Fig5_1.6T transceiver}e, at the wavelength of 1562.2 nm, the hybrid scheme only requires 29-tap FFE to arrive below the HD-FEC threshold, whereas the DSP-only scheme requires a 965-tap FFE and only achieves SD-FEC. Since most of the computing is offloaded to the optical domain, even as the number of wavelengths and data rate per wavelength scale in next-generation DCIs, lightweight DSP can remain efficient and feasible to implement.

\subsection{Latency and power consumption analysis}\label{subsec2}

\begin{figure}[htbp] 
    \centering 
    \includegraphics[width=1\textwidth]{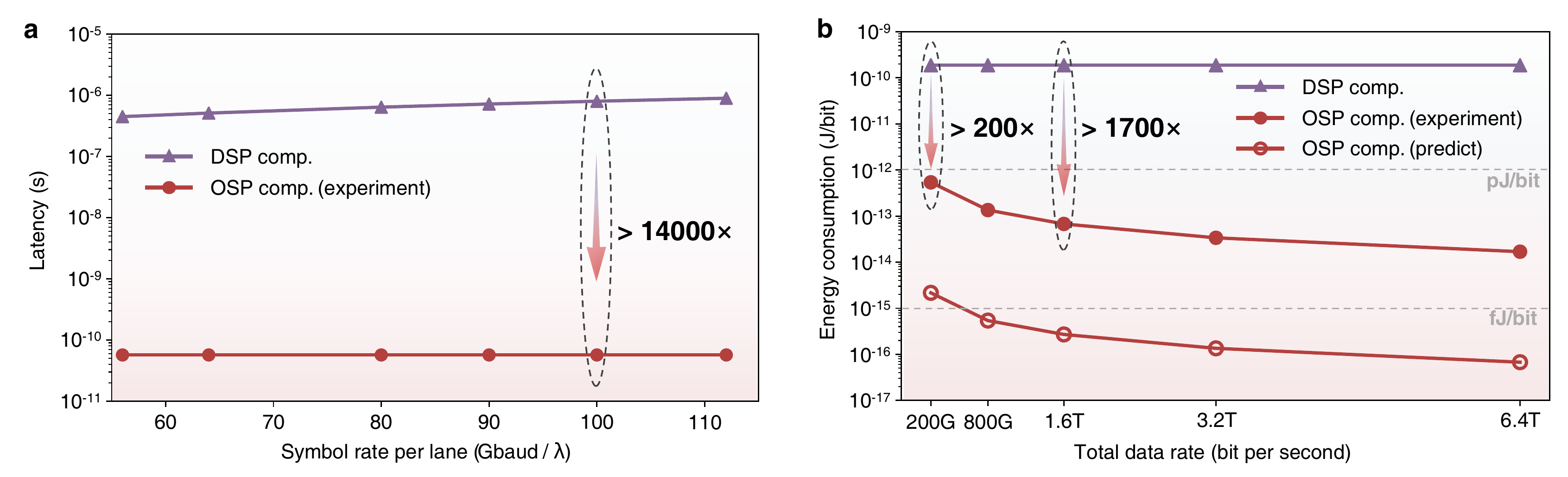} 
    \caption{Latency and energy consumption comparison. (a) Latency of single channel (1550 nm) for PAM4 signals with different symbol rates over a 5 km fiber link with the DSP and OSP compensation. (b) Energy consumption of OSP and DSP compensation for scaling 200 Gbps/$\lambda$ with more channels (compensation for 5 km SMF in C-band, equivalent to over 40 km in O-band).} 
    \label{fig:Fig6_Energy and power} 
\end{figure}

Our OSP not only surpasses DSP in signal compensation performance but also drastically reduces processing latency and power consumption. The processing latency of our OSP system is primarily determined by the signal propagation time in the OSP. With a total waveguide length of approximately 4,000 µm, the corresponding propagation time is estimated to be 57 ps. A key distinction between OSP and DSP is that the latency in DSP is fundamentally constrained by the system clock, $f_{clock}$. In most DSP systems, $f_{clock}$ is typically set at a few hundred MHz for optimal energy efficiency. Estimating the exact latency in a DSP chip is challenging, as it depends not only on the required operations but also on the specific ASIC design. To simplify the problem, we assume the best-case scenario for DSP, where DSP chips operate at the line rate through parallelization, leaving only the delay in serial-to-parallel (s/p) conversion to consider. The latency of s/p conversion is given by $2B_{Data}/f_{clock}^2$, assuming 2 samples per symbol~\cite{huang22JLT}. Even in this over-optimistic scenario, the minimum latency in DSP is 0.8 µs for 100 Gbaud PAM4 signals, which is over 14,000 times larger than the 57 ps latency of our OSP, as illustrated in Fig.~\ref{fig:Fig6_Energy and power}a. In addition, the latency of the OSP remains constant as the data rate increases, whereas in DSP, the latency increases at least linearly with the data rate. 

The power consumption of OSP mainly comes from the power consumption of microheaters. With optimized currents, the power consumption to process 100 Gbaud PAM4 over 5 km C-band transmission is approximately 108 mW, leading to an energy efficiency of 0.54 pJ/bit. When estimating the energy consumption of DSP, we assume advanced 3 nm CMOS technology. For simplicity, we exclude the power consumption of the serializer/deserializer, which is a highly challenging and power-consuming module in DSP, particularly for high-speed signals. We only account for basic multiplication and addition operations in DSP algorithms. The energy consumption of multiplication $E_{op,M}$ and addition $E_{op,A}$ are 208.5 fJ/b and 26 fJ/b, according to Pillai et al~\cite{pillai14JLT}. The estimated power consumption of FFE with 800 taps for 100 Gbaud/$\lambda$ PAM4 signal is at least 37.5 W, which far exceeds the thermal management capacity of current optical transceivers~\cite{qsfpdd2021}. Therefore, our OSP can achieve over 200-fold reduction in power consumption compared to DSP in the single channel scenario. Furthermore, thanks to its inherently and highly parallel processing capacity, the OSP can achieve even better energy consumption at higher data rates, as shown in Fig.~\ref{fig:Fig6_Energy and power}b. The energy efficiency can be improved to 67.5 fJ/bit in 1.6 T (200 Gbps × 8) DCIs, with projections reaching sub-fJ/bit levels through the adoption of more power-efficient phase shifters such as barium titanate (BTO)~\cite{li23JLT}, micro-electron-mechanical systems (MEMS)~\cite{kim23NP}, and low-loss integrated platforms. (See Supplementary Note 4 and 5 for a detailed calculation).

The scaling law for OSP, in terms of both latency and energy efficiency, makes it extremely attractive for next-generation high-speed optical interconnects, particularly in hyperscale data centers.

\section{Discussion and conclusions}\label{sec3}
One common challenge of OSP is the chip insertion loss, which results from on-chip interference effects and insertion losses in the waveguides and components, leading to a total insertion loss of 15 dB. Specifically, the fiber-to-chip loss of the OSP chip is around 6 dB per edge coupler. However, this loss can be mitigated by integrating the photodetector with the OSP, eliminating the need for additional edge couplers. In the future, such insertion loss can be compensated using on-chip optical amplifiers such as III-V/Si semiconductor optical amplifiers~\cite{7516667}. 

In conclusion, we present an experimental demonstration of an all-optical integrated OSP for data center interconnects, showcasing DSP-free and real-time signal processing of up to 112 Gbaud/$\lambda$ PAM4 over a 5 km C-band optical fiber link. Additionally, we demonstrate the scalability of the OSP to WDM operations, supporting 8-channel 1.6T transmission with a single optical chip. This processing capability significantly surpasses the performance of most advanced DSP chips, algorithms, and photonic processors. Furthermore, the OSP offers a remarkable energy efficiency of 67.5 fJ/bit and a latency of 57 ps, resulting in a 1,700-fold reduction in energy consumption and a 14,000-fold decrease in latency compared to advanced DSP chips at 1.6T transmission. By leveraging MEMS~\cite{kim23NP} or nonvolatile materials such as BTO~\cite{li23JLT} for weight programming, this enhancement would further improve the energy efficiency to 0.27 fJ/bit. The gap in advantages continues to scale with higher data rates, positioning it as a transformative alternative for next-generation DCIs to meet the surging data traffic driven by AI advancements.

\section{Methods}\label{sec11}

\subsection{Design, fabrication, and package of the OSP chip}

We fabricated our device using a standard silicon-on-insulator (SOI) fabrication process at a silicon photonic foundry. This platform also includes an 800 nm thick oxide passivation layer, a Ti/W heating filament layer, and an aluminum (Al) routing layer. The OSP chip consists of fully etched waveguides with a width of 500 nm and a thickness of 220 nm, allowing for single-mode operation on different polarizations at 1550 nm. The input and output edge couplers are nanotapers with a 2.5 µm optical field radius and fix the polarization to the transverse electric (TE) mode. The OSP chip consists of three cascaded time-delay reservoirs and an eight-channel photonic readout. The time-delay reservoirs have delay lengths of 1270, 2540, and 640 µm, corresponding to time delays of around 18, 36, and 9 ps, respectively. Different channels of the readout have a delay interval of 350 µm, with respect to round 5 ps. The delay waveguides here are designed in a square shape, instead of the spiral waveguide, to minimize excessive radiation loss. Microheaters on the waveguide allow for flexible control of programmable parameters, accommodating input signals with varying wavelengths or power. The heaters have a width of 3 µm and a length of 220 µm, with a resistance of about 300 $\Omega$. The power consumption for a $\pi$ phase tuning operation is approximately 25 mW. Routing traces are deposited to connect the microheaters to electrical metal pads. Deep trenches are positioned beside vertically adjacent microheaters to suppress the thermal crosstalk. Additionally, we wire-bonded the chip to a designated printed circuit board (PCB) to ensure stable and convenient access to all on-chip phase shifters and amplitude controllers. We have designed the PCB with a hollowed-out structure at the center, specifically to house the chip. The thermistor and TEC module are placed close to and under the chip separately, to sense and control the chip temperature. The bottom features an aluminum casing, enabling direct contact between the chip and the metal for efficient heat conduction. This arrangement allows for optimal and stable temperature control.

\subsection{Experimental setup for IMDD data transmission}

More details are presented here for the experiment setup shown in Fig.~\ref{fig:Fig2_Literature}a. For the C-band IMDD data transmission, the TLS (Keysight N8844C) is tuned to around 1550 nm. The communication signals are generated by an AWG (Keysight, M8199A) at 256 GSa/s, driven by an electrical amplifier (SHF 804C), and modulated by a 40-GHz commercial TFLN intensity modulator (LIOBATE LN-40). The first EDFA amplifies the optical signal up to 20 dBm and then the first VOA controls optical power launched into the fiber link made of a standard G.652D standard single-mode fiber with a C-band nominal loss coefficient of 0.2 dB/km. After 5-km fiber transmission, the communication signals are degraded by severe chromatic dispersion in the fiber and nonlinearity existing in the experiment setup. A polarization controller allows the tuning of the local compression and torque applied to the fiber itself, inducing polarization change to match our photonic chip.  The photonic chip is programmed by a multi-channel source-meter unit (National Instruments PXIe 4163). Through optimizing the on-chip parameters (i.e., currents on heaters), the signal distortions can be learned and effectively equalized, directly in the optical domain. The output signal of the photonic chip is sent to the second EDFA with a small signal gain of about 30 dB followed by a second VOA to control the received optical power. A tunable optical filter with 100 GHz bandwidth cleans up the signal from the out-of-band amplified spontaneous emission noise added by the amplification stages. The filtered signal was directly detected by a commercial 70-GHz photodetector (COHERENT XPDV3120R). The received electrical signal was finally recorded by a real-time oscilloscope (RTO, Keysight UXR0592AP) at 256 GSa/s. A time-domain small-tap feed-forward equalizer (FFE) is employed to compensate for BtB impairments, mainly caused by bandwidth limitations of the modulator and the photodiode. When the OSP is optimized for fiber distortion compensation, the FFE taps are kept identical to those in the BtB configuration, and optical power before the photodetector is maintained constantly across experiments to ensure consistency.

\backmatter
\bmhead{Funding}
This work was supported by ITF ITS/237/22, ITS/226/21FP, RGC ECS 24203724, NSFC 62405258, RGC YCRG C4004-24Y, C1002-22Y, RNE-p4-22 of the Shun Hing Institute of Advanced Engineering, NSFC/RGC Joint Research Scheme N CUHK444/22.

\bmhead{Acknowledgments}
We acknowledge Applied Nanotools for chip fabrication and NOEIC for chip package.

\bmhead{Author contributions}
B.W. and C.H. conceived the idea. B.W. designed, simulated, and characterized the photonic chip. B.W. performed the high-speed signal communication experiments. B.W. conducted offline DSP with the help of Q.X. B.W. and C.H. analyzed the data with the help of Q.X., T.X., L.F., and S.L. B.W. and C.H. prepared the manuscript with the support of all authors. J.D. and J.Z. provided suggestions and feedback during the revisions. C.H. supervised the project. All the authors discussed the data and contributed to the manuscript.

\bmhead{Conflicts of interest}
The authors declare no conflicts of interest.

\bmhead{Data availability}
Data underlying the results presented in this paper are not publicly available at this time but may be obtained from the authors upon reasonable request.

\bmhead{Code availability}
The codes in this study are available from the corresponding authors upon request.

\bmhead{Supplementary information}

Supplementary Note.









\begin{appendices}






\end{appendices}


\bibliography{sn-bibliography}

\end{document}